\documentclass[a4paper]{spie}
\usepackage{graphicx}
\usepackage{amsfonts,amssymb,amsbsy,amsmath}

\usepackage{colortbl}

\title{Photon momentum and optical forces in cavities} 
\author{Mikko Partanen,$^1$ Teppo H\"ayrynen,$^{1,2}$ Jani Oksanen,$^1$ and Jukka Tulkki$^1$\skiplinehalf
$^1$Engineered Nanosystems group, School of Science\\Aalto University, P.O. Box 12200, 00076 Aalto, Finland\\
$^2$DTU Fotonik, Department of Photonics Engineering, Technical University of
Denmark, \O rsteds Plads, Building 343, DK-2800 Kongens Lyngby, Denmark
}

\begin{document} 

\maketitle 

\begin{abstract}
During the past century, the electromagnetic field momentum in
material media has been under debate in the Abraham-Minkowski controversy
as convincing arguments have been advanced in favor of both
the Abraham and Minkowski forms of photon momentum.
Here we study the photon momentum and optical
forces in cavity structures in the cases of dynamical and steady-state fields.
In the description of the single-photon transmission process,
we use a field-kinetic one-photon theory. Our model
suggests that in the medium photons couple with the
induced atomic dipoles forming polariton quasiparticles with the Minkowski
form momentum. The Abraham momentum can be associated to the electromagnetic
field part of the coupled polariton state.
The polariton with the Minkowski
momentum is shown to obey the uniform center of mass of energy motion
that has previously been interpreted to support only
the Abraham momentum.
When describing the steady-state nonequilibrium field distributions we use the recently
developed quantized fluctuational electrodynamics (QFED) formalism.
While allowing detailed studies of light propagation
and quantum field fluctuations in
interfering structures, our methods also provide practical tools for modeling
optical energy transfer and the formation of thermal balance in nanodevices
as well as studying electromagnetic forces in optomechanical devices. 
\end{abstract}

\section{Introduction}

Investigations of radiation pressure and the momentum of
light in dielectrics have frequently involved arguments about the
correct form of the electromagnetic field momentum in material media
\cite{Cho2010,Leonhardt2006,Barnett2010b,Barnett2010a,Leonhardt2014}.
The Abraham and Minkowski forms for the single photon momentum are given by
$\hbar k_0/n$ and $\hbar k_0n$, which naturally
depend on the vacuum wavenumber $k_0$ but also
introduce contradicting and confusing dependencies on refractive
index $n$. During the past century,
powerful arguments have been advanced in favor of both momenta
\cite{Barnett2010b,Barnett2010a}
and various experimental setups measuring the forces
due to light also seem to support both momenta
\cite{Campbell2005,Sapiro2009,Jones1954,Jones1978,Walker1975,She2008,Zhang2015}.

In this work, we investigate the propagating field momentum and optical forces
in cavity structures. The light pulses and single photon fields are described by
using a field-kinetic theory that is based on the
covariance principle \cite{Schwartz2014}, which
states that the laws of physics are the same for all inertial
observers. The steady-state field distributions
are instead described by using the quantized fluctuational electrodynamics
(QFED) formalism \cite{Partanen2014a,Partanen2014c,Partanen2015a}.
The QFED approach for formulating the field operators is based on
defining position-dependent
photon ladder operators that obey canonical commutation
relations \cite{Partanen2014a}.
Our methods allow detailed studies of quantum field fluctuations
in interfering structures, but also provide
practical tools for modeling optical energy transfer and the formation
of thermal balance in nanodevices as well as
studying electromagnetic forces in optomechanical devices.

The manuscript is organized as follows: The field-kinetic model
of photon propagation is presented in Sec.~\ref{sec:fieldkineticmodel}.
Section \ref{sec:qfed} covers the principles of calculating
electromagnetic forces in cavity structures by using the QFED model.
Finally, conclusions are drawn in Sec.~\ref{sec:conclusions}.

\begin{figure}
\begin{center}
\includegraphics[width=0.8\textwidth]{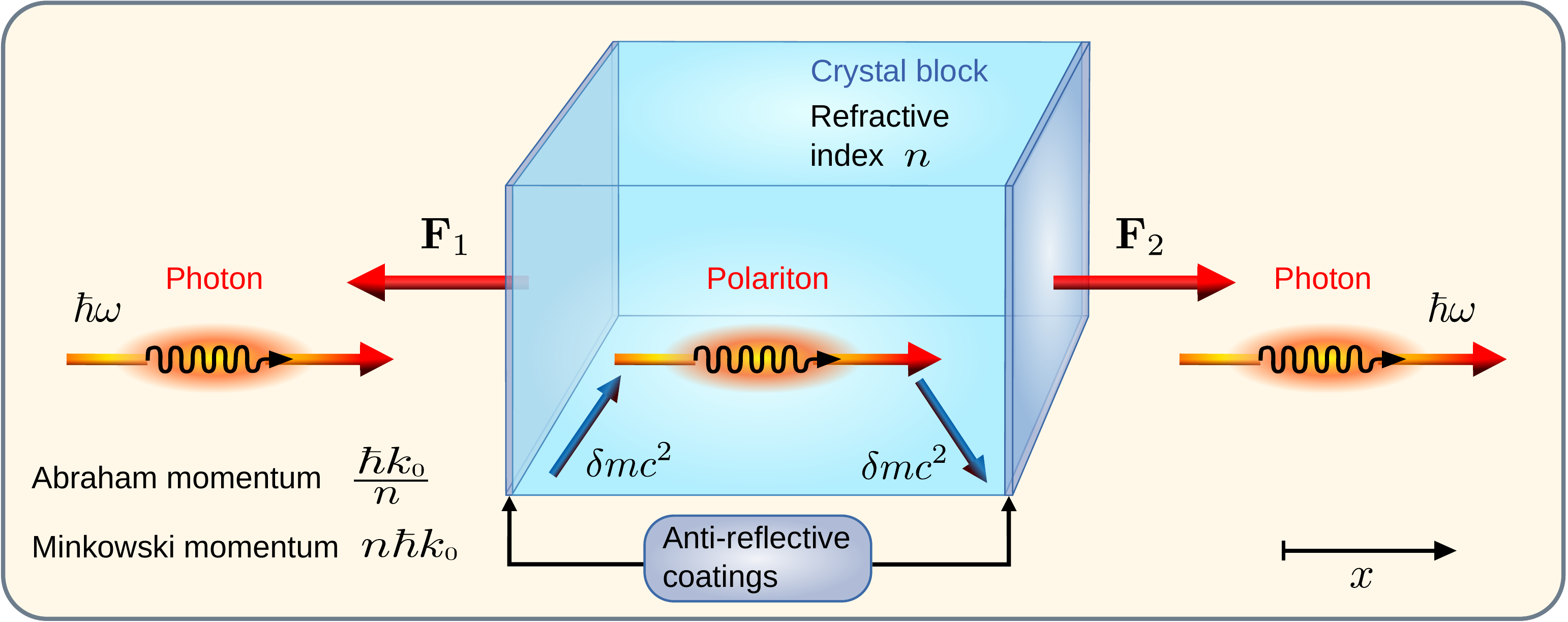}
\caption{\label{fig:problem}(Color online) Schematic illustration of a single photon transmission through
a crystal block with refractive index $n$. The photon is first incident from vacuum.
When entering the medium it couples to the induced electric dipoles in the medium forming a polariton
quasiparticle. At the second end of the crystal the photon continues to propagate
in vacuum. At the photon entrance and exit the medium block
experiences forces $\mathbf{F}_1$ and $\mathbf{F}_2$.}
\end{center}
\end{figure}

\section{Field-kinetic theory of photon propagation}
\label{sec:fieldkineticmodel}

In this section, we present theoretical observations
regarding the single photon transmission
through a crystal block with refractive index $n$ illustrated in
Fig.~\ref{fig:problem}(a) by using a field-kinetic
one-photon model. In our theory, we generalize the Feynman's
description of light propagating in solids
\cite{Feynman1964} where the light quantum interacts
with the induced atomic dipoles in the medium and forms a coupled state of
light and matter that we call a polariton.

Here we use the concept of the polariton in a meaning that
differs from its conventional use in the context of
the phonon-polariton and the exciton-polariton quasiparticles.
In these conventional cases the total energy of the polariton
oscillates between two different eigenstates that
represent physically different expressions of
the total polariton energy. As discussed above,
in our case the polariton
is sooner a coupled state of electrons, ions,
and the electromagnetic
wave and it behaves like a Bloch state of electrons in solids.
It propagates through the crystal without scattering
preserving its energy and momentum and also
conveys with itself a small but fixed amount
of rest mass. It is also
vital that the photon energy
is far from the energy of the elementary electronic or ionic
excitations of the medium so that absorption
and spontaneous emission processes do not occur.

In the presentation of the field-kinetic one-photon model we
first consider the energy and momentum
conservation laws and the related covariance condition.
The energy-momentum covariance only gives the relation
between the energy and momentum, but does not
uniquely determine the polariton momentum in the medium.
Therefore, we also consider further physical conditions
which can be used to determine the
polariton momentum.

\subsection{Energy and momentum conservation}
We consider the total energy-momentum
four-vectors of different parts of the composite system.
These vectors consist of energy and momentum as
$(E/c,p_x,p_y,p_z)$ \cite{Schwartz2014}.
The initial total four-momentum is written as a sum of the free photon
four-momentum $P_0=(\hbar k_0,\hbar k_0,0,0)$ and the
four-momentum $P_M=(Mc,0,0,0)$ for the medium block at rest as $P_\mathrm{tot}=P_0+P_M$.

In the field-kinetic description of the polariton,
we assume that the induced dipoles of the medium
carry a small but finite rest mass $\delta m$ that will be determined
from the conservation laws and the covariance conditions.
The total polariton energy is then given by the sum
of the initial electromagnetic energy $E_\mathrm{f}=\hbar\omega$
and the rest energy $E_\mathrm{d}=\delta mc^2$ corresponding to $\delta m$
as $E=\hbar\omega+\delta mc^2$. In the medium, the total polariton
energy $E$ propagates with velocity $v=c/n$.
The total momentum of the polariton denoted by $p$ is a sum of the field and
dipoles related contributions $p_\mathrm{f}$ and
$p_\mathrm{d}$ and it will be uniquely determined in Sec.~\ref{sec:polaritonmomentum}.

When the photon enters the medium, the photon couples with the
atoms in the medium and the total energy and momentum
of the system are shared by the propagating polariton
$P_\mathrm{pol}=(E/c,p,0,0)$ and the recoiling medium block
$P_\mathrm{med}=(M_\mathrm{r}c,M_\mathrm{r}V_\mathrm{r},0,0)$.
Here $M_\mathrm{r}=M-\delta m$ is the recoil mass
of the medium block and $V_\mathrm{r}$ is the recoil velocity
in the $x$-direction in Fig.~\ref{fig:problem}.
The total four-momentum must be conserved and thus we have
$P_\mathrm{tot}=P_\mathrm{pol}+P_\mathrm{med}$.

The unknown quantities of the model can be uniquely
solved for a given total polariton momentum $p$
by applying the energy and momentum conservation laws and the
energy-momentum covariance condition $E^2/c^2-p_x^2-p_y^2-p_z^2=m_0^2c^2$,
where $m_0$ is the effective rest mass \cite{Schwartz2014}.
The conservation of energy corresponds to the conservation
of the first component of the four-momentum and it is written as
\begin{equation}
\hbar\omega+Mc^2=E+M_\mathrm{r}c^2.
\label{eq:energyconservation}
\end{equation}
The momentum conservation instead corresponds to the conservation of the other components of
the four-momentum and it is given for the nonzero second component by
\begin{equation}
\hbar k_0=p+M_\mathrm{r}V_\mathrm{r}.
\label{eq:momentumconservation}
\end{equation}
The principle of covariance also requires that the four-momenta
obey the energy-momentum covariance condition.
The covariance-condition-obeying energy $E=\gamma m_0c^2$ and
momentum $p=\gamma m_0v$, where
$m_0$ is the effective rest mass and
$\gamma=1/\sqrt{1-v^2/c^2}$ is the Lorentz factor, obey $E=pc^2/v$.
The covariance condition therefore directly relates
the corresponding momenta $p_\mathrm{f}=E_\mathrm{f}v/c^2$ and
$p_\mathrm{d}=E_\mathrm{d}v/c^2$ to the
electromagnetic field and dipoles related energies $E_\mathrm{f}=\hbar\omega$
and $E_\mathrm{d}=\delta mc^2$.

By applying the conservation laws and the covariance condition,
the induced dipoles related mass $\delta m$ and the
medium block recoil velocity $V_\mathrm{r}$ can be uniquely
determined for a given total polariton momentum $p$ as
$\delta m=np/c-\hbar\omega/c^2$ and
$V_\mathrm{r}=(\hbar\omega-cp)/(M_\mathrm{r}c)$.
The energy and momentum contributions
are presented in Table \ref{tbl:table} for the
general, Abraham, and Minkowski form polariton momenta $p$.
It can be seen that, in the case of the Abraham momentum,
the dipoles related quantities are zero, whereas, in the case of the
Minkowski momentum, the total polariton quantities include dipoles
related parts that make the total polariton momentum to be given
by the Minkowski form.

\begin{table}
 \centering
 \renewcommand{\arraystretch}{1.2}
 \caption{\label{tbl:table}
 Polariton model energies and momenta calculated by using
 the general, Abraham, and Minkowski form polariton momenta $p$.
 Here $E=E_\mathrm{f}+E_\mathrm{d}$ and $p=p_\mathrm{f}+p_\mathrm{d}$
 are the total energy and momentum of the polariton and the
 quantities with subscripts f and d are, respectively, related to the
 electromagnetic field and the induced dipoles.
 }
 \vspace{0.1cm}
 \begin{tabular}{|c|c|c|c|}
   \hline
   & General & Abraham  & Minkowski\\[2pt]
   \hline
   $E$ & $npc$ & $\hbar\omega$ & $n^2\hbar\omega$\\[2pt]
   $E_\mathrm{f}$ & $\hbar\omega$ & $\hbar\omega$ & $\hbar\omega$\\[2pt]
   $E_\mathrm{d}$ & $npc-\hbar\omega$ & $0$ & $(n^2-1)\hbar\omega$\\[2pt]
   $p$ & $p$ & $\hbar k_0/n$ & $n\hbar k_0$\\[2pt]
   $p_\mathrm{f}$ & $\hbar k_0/n$ & $\hbar k_0/n$ & $\hbar k_0/n$\\[2pt]
   $p_\mathrm{d}$ & $p-\hbar k_0/n$ & $0$ & $(n-\frac{1}{n})\hbar k_0$\\[2pt]
   \hline
 \end{tabular}
\end{table}

Isolated systems like the photon plus the medium block in Fig.~\ref{fig:problem}
are known to obey uniform motion described by a constant center of mass of energy velocity (CEV).
According to our results, the CEV
is written for the isolated system of a medium block and the photon
before and after the photon has entered the medium as
\begin{equation}
 V_\mathrm{CEV}=\frac{\sum_iE_iv_i}{\sum_iE_i}=\frac{\hbar\omega c}{\hbar\omega+Mc^2}=\frac{Ev+M_\mathrm{r}c^2V_\mathrm{r}}{E+M_\mathrm{r}c^2}.
 \label{eq:uniformmotion2}
\end{equation}
The equality of the numerators is nothing else than
the conservation of momentum in Eq.~\eqref{eq:momentumconservation}
and the equality of the denominators corresponds to the
energy conservation in Eq.~\eqref{eq:energyconservation}.
The above calculations are independent of the exact form of the polariton
momentum $p$, which essentially shows that a covariant
theory obeying the constant CEV motion
can be formulated for both the Abraham and Minkowski momenta.
This is essential as before only the
Abraham momentum has been reasoned to obey
the constant CEV motion \cite{Barnett2010b,Barnett2010a}.
However, as described below, there exist further conditions that may
uniquely define the polariton momentum.

\subsection{Determination of the polariton momentum}
\label{sec:polaritonmomentum}

As the above covariant theory does not uniquely define
the exact form of the polariton momentum $p$,
we next consider further physical conditions which
can be used to directly determine the polariton momentum.
One approach to determine the polariton momentum
is given by the polariton Bloch state concept in which the polariton
is considered to be a coupled Bloch state of light and matter.
We follow the theory of the electronic structure of solids
and suggest that the wavefunction of the polariton
Bloch state, which can propagate through the medium without
scattering or absorption, must be of the form
$e^{i\mathbf{k}\cdot\mathbf{r}}u_{\mathbf{k}}(\tau)$,
where $\mathbf{r}$ is the position vector,
$\mathbf{k}$ is the wavevector of the polariton,
and $u_{\mathbf{k}}(\tau)$ is the amplitude function
which includes the microscopic structure of matter and its
coupling with light through the generalized coordinate $\tau$.
The quantum mechanical momentum operator is given by
$\hat{\mathbf{p}}=-i\hbar\nabla$, where $\nabla$ is the vector differential
operator. The momentum expectation
value then obviously becomes $\langle\hat{p}\rangle=\hbar k$.
Inside a medium, the wavelength of light decreases to
$\lambda=\lambda_0/n$, where $\lambda_0$ is the wavelength
in vacuum. Therefore, from the momentum
expectation value, it directly follows that
$\langle\hat{p}\rangle=\hbar k=2\pi\hbar/\lambda=n2\pi\hbar/\lambda_0=n\hbar k_0$
and thus the Minkowski momentum is a natural consequence of the polariton
Bloch state concept.
This will be discussed further in the forthcoming manuscript
where we also use the Lorentz force law and the Maxwell's equations
to show that this simplified model is indeed fully consistent with the
semiclassical continuum picture and
the energy-momentum tensors of light and matter.

\subsection{Physical consequences}
One of the most evident physical consequences
of our analysis is that energy corresponding to
the rest mass $\delta m$ is effectively attached to the photon
when the photon propagates inside the medium as a polariton quasiparticle.
Note, however, that the nonzero rest mass is a property of the polariton not the photon.
As we obtained the Minkowski momentum for the polariton, we have
$\delta m=(n^2-1)\hbar\omega/c^2>0$,
which essentially means that the polariton transfers the mass $\delta m$
from the first to the second interface of the block
and the medium is left in a nonequilibrium
state which later on returns to equilibrium through relaxation processes.
Since the photon energy is conserved in the transmission process,
the energy of this nonequilibrium state is very
close to the energy of the initial state and the relaxation
processes are practically elastic.

As an example of the Minkowski form polariton momentum for $\hbar\omega=1$ eV
and $n=2$, we have $\delta mc^2=3$ eV.
If the mass density of the medium is 1000 kg/m$^3$, this mass transfer effectively
corresponds to a displacement of a medium cube with side length $2\times 10^{-13}$ m,
which corresponds to a small fraction of an atom. This displacement
is divided into a region of many atoms and one can well expect
that the energy required to produce such a small displacement of
atoms along the path of the photon is meaningless
in a dispersionless medium and the photon
does not lose energy in the transmission process.
One can also respectively speculate that in dispersive media the atomic
displacements may be inelastic when the photon
loses energy which leads to dispersion.

\begin{figure}
\centering
\includegraphics[width=0.78\textwidth]{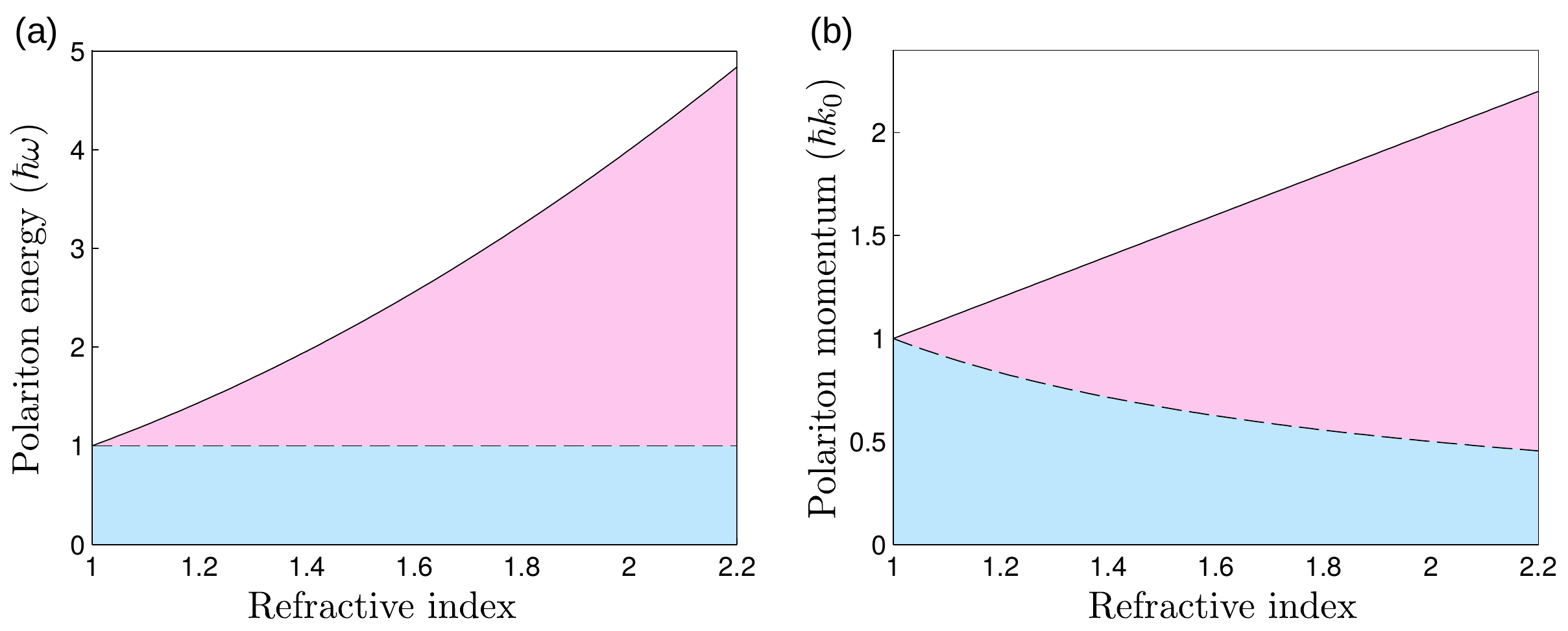}
\caption{\label{fig:polariton}(Color online) (a) Polariton energy and (b)
polariton momentum as a function of the refractive index. The solid
lines denote the total polariton energy $E=E_\mathrm{f}+E_\mathrm{d}$
and momentum $p=p_\mathrm{f}+p_\mathrm{d}$ and the dashed lines
denote the electromagnetic field associated parts $E_\mathrm{f}=\hbar\omega$ and
$p_\mathrm{f}=\hbar k_0/n$ of the
corresponding quantities. The differences of the solid and dashed lines
correspond to the energy and momentum parts $E_\mathrm{d}=\delta mc^2=(n^2-1)\hbar\omega$
and $p_\mathrm{d}=(n-\frac{1}{n})\hbar k_0$ carried by the induced electric dipoles in the medium.}
\end{figure}

Our results suggest that the experimental measurements that
directly measure the polariton momentum or the force due to a light
beam inside a medium \cite{Jones1954,Jones1978,Campbell2005,Sapiro2009}
must give the Minkowski momentum for the polariton.
The Abraham momentum seems to be only supported by indirect measurements
\cite{Walker1975,She2008,Zhang2015} and
theoretical arguments that do not take the polariton associated
rest mass and the related mass transfer and its relaxation into account.

The uniquely defined polariton model quantities corresponding
to the Minkowski form polariton presented in the last
column in Table \ref{tbl:table} are illustrated in Fig.~\ref{fig:polariton}.
Figure \ref{fig:polariton}(a) shows the total polariton energy
and its contributions associated to the electromagnetic field
and the induced electric dipoles in the medium as
a function of the refractive index.
The total polariton energy in Fig.~\ref{fig:polariton}(a)
increases as a function of the refractive index
due to the increasing polariton associated rest
mass. The increased rest mass is also
related to the reduction of the propagation velocity of light
in the medium. The energy contribution associated
to the electromagnetic field part of the coupled polariton
state remains constant $\hbar\omega$.

The polariton momentum is presented as a function of the
refractive index in Fig.~\ref{fig:polariton}(b).
The Minkowski form total polariton momentum increases linearly with the
increasing refractive index. Its electromagnetic
field contribution given by the Abraham momentum
instead decreases with the increasing refractive index.
The difference of the Minkowski and Abraham momenta
is carried by the induced electric dipoles
in the medium.

\section{Quantized fluctuational electrodynamics}
\label{sec:qfed}

To provide additional insight on the forces present at the interfaces
in Fig.~\ref{fig:problem}, we briefly review the methods to calculate
the forces $\mathbf{F}_1$ and $\mathbf{F}_2$ of the figure by using
the QFED formalism \cite{Partanen2014a,Partanen2014c,Partanen2015a,Partanen2014b}.
Fully equivalent steady-state forces can also be obtained directly from the
corresponding Maxwell's stress tensor.
Since, in the QFED formalism, we study
time-independent steady-state fields, we do not face the problem of defining
the photon momentum and the Abraham-Minkowski controversy. 
In the QFED, the forces are calculated by using the operator
form of the classical Maxwell's stress tensor. It follows
that the $x$-component of the spectral force density expectation
value is given by \cite{Partanen2014c}
\begin{align}
 \langle\hat{\mathcal{F}}_x(x,t)\rangle_\omega & =-\frac{\hbar\omega}{2}\Big(\frac{\partial}{\partial x}\rho(x,\omega)\Big)
  -\hbar\omega\Big(\frac{\partial}{\partial x}\rho(x,\omega)\Big)\langle\hat n(x,\omega)\rangle
-\hbar\omega\rho(x,\omega)\frac{\partial}{\partial x}\langle\hat n(x,\omega)\rangle,
 \label{eq:forcedensity}
\end{align}
where $\rho(x,\omega)$ is the electromagnetic local density of states (LDOS) and
$\langle\hat n(x,\omega)\rangle$ is the position-dependent photon number expectation value.
The first term in Eq.~\eqref{eq:forcedensity} corresponds
to the familiar zero-point Casimir force (ZCF) \cite{Rodriguez2011,Antezza2008},
the second term is known as the thermal Casimir force (TCF) \cite{Passante2007,Sushkov2011,Klimchitskaya2008},
and the last term arising from the changes in the total photon number is called
a nonequilibrium Casimir force (NCF) \cite{Partanen2014c} since it disappears at thermal
equilibrium when the derivative of the photon number is zero.
The net force on an area $S$ of a solid object extending from $x_1$ to $x_2$ can then be obtained
by integrating the force density in Eq.~\eqref{eq:forcedensity}
as $\langle\hat{F}(t)\rangle_\omega=S\int_{x_1}^{x_2}\langle\hat{\mathcal{F}}_x(x,t)\rangle_\omega dx$
\cite{Partanen2014c}. Equivalently, the net force can be also obtained
by using the concept of electromagnetic pressure.
The electromagnetic pressure along the $x$ direction is given by \cite{Partanen2014c}
 \begin{equation}
  \langle\hat{\mathcal{P}}(x,t)\rangle_\omega=\hbar\omega\rho(x,\omega)\Big(\langle\hat n(x,\omega)\rangle+\frac{1}{2}\Big).
  \label{eq:pressure}
 \end{equation}
Therefore, the net force on an object extending from $x_1$ to $x_2$
can be obtained as $\langle\hat{F}(t)\rangle_\omega=S[\langle\hat{\mathcal{P}}(x_1,t)\rangle_\omega-\langle\hat{\mathcal{P}}(x_2,t)\rangle_\omega]$
\cite{Partanen2014c}.

\begin{figure}
\centering
\includegraphics[width=0.48\textwidth]{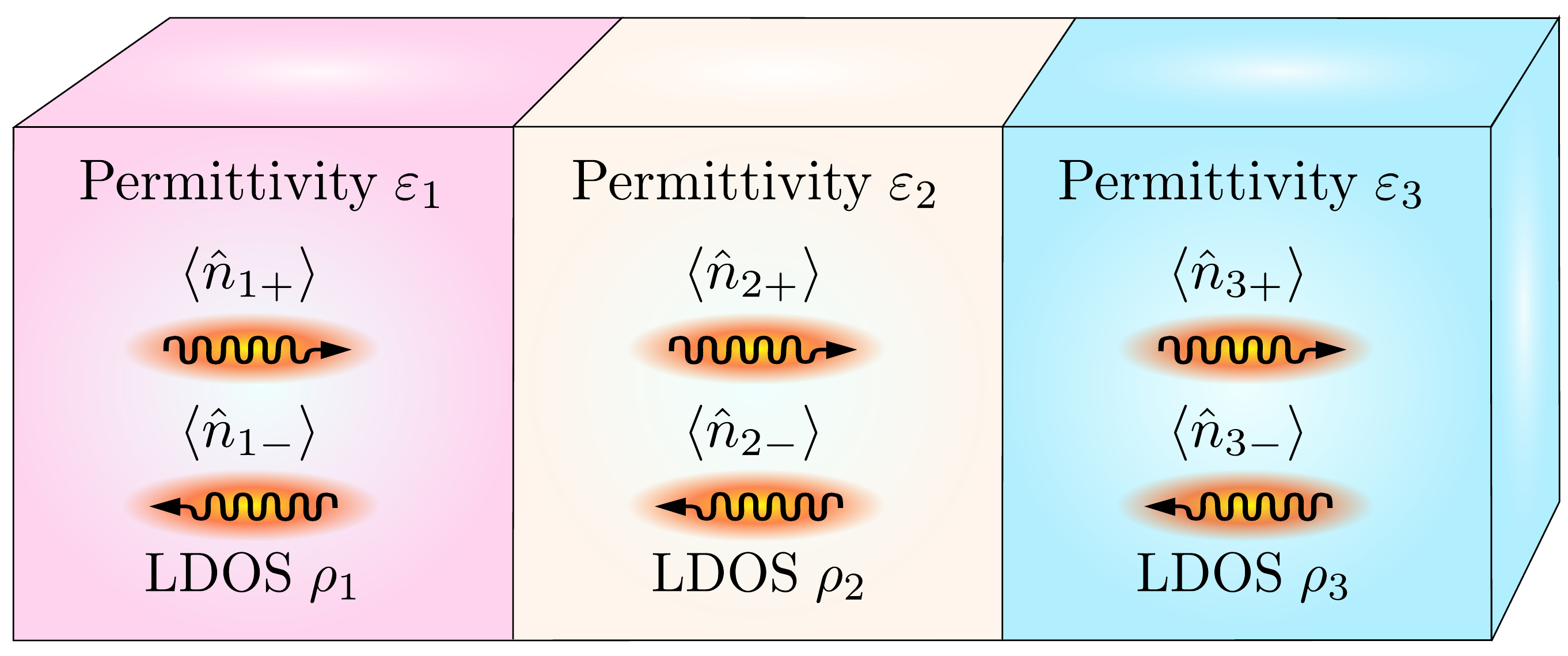}
\caption{\label{fig:cavity}(Color online) Optical cavity structure consisting of three homogeneous layers.
The left and right propagating field photon-number expectation values and LDOSs
are presented in each layer.}
\end{figure}

The photon numbers propagating to the left and right in different
parts of the cavity geometry corresponding to the medium block
in Fig.~\ref{fig:problem} are illustrated in Fig.~\ref{fig:cavity}.
As there are no losses, the left and right propagating and total field photon numbers are piecewise
continuous and only depend on the cavity geometry and the input fields $\langle\hat n_{1+}\rangle$
and $\langle\hat n_{3-}\rangle$ incident from the left and right.
The refractive indices of the three media are given by
$\sqrt{\varepsilon_1}$, $\sqrt{\varepsilon_2}$, and $\sqrt{\varepsilon_3}$.
In different regions of the geometry, the propagating photon numbers
are written as \cite{Partanen2015a}
\begin{align}
 \langle\hat n_{1-}\rangle & =\textstyle|\mathcal{R}_1|^2\langle\hat n_{1+}\rangle+\sqrt{\varepsilon_1/\varepsilon_3}\,|\mathcal{T}_1'\mathcal{T}_2'|^2\langle\hat n_{3-}\rangle,\nonumber\\
 \langle\hat n_{2+}\rangle & =\frac{\sqrt{\varepsilon_2/\varepsilon_1}\,|\mathcal{T}_1|^2\langle\hat n_{1+}\rangle+\sqrt{\varepsilon_2/\varepsilon_3}\,|\mathcal{T}_2'\mathcal{R}_1'|^2\langle\hat n_{3-}\rangle}{\mathrm{Re}[1+2\mathcal{R}_1'\mathcal{R}_2\nu_2e^{2ik_2d_2}]},\nonumber\\
 \langle\hat n_{2-}\rangle & =\frac{\sqrt{\varepsilon_2/\varepsilon_1}\,|\mathcal{T}_1\mathcal{R}_2|^2\langle\hat n_{1+}\rangle+\sqrt{\varepsilon_2/\varepsilon_3}\,|\mathcal{T}_2'|^2\langle\hat n_{3-}\rangle}{\mathrm{Re}[1+2\mathcal{R}_1'\mathcal{R}_2\nu_2e^{2ik_2d_2}]},\nonumber\\
 \langle\hat n_{3+}\rangle & =\textstyle\sqrt{\varepsilon_3/\varepsilon_1}\,|\mathcal{T}_1\mathcal{T}_2|^2\langle\hat n_{1+}\rangle+\textstyle|\mathcal{R}_2'|^2\langle\hat n_{3-}\rangle,
 \label{eq:photons}
\end{align}
where $d_2$ is the width of the cavity, $k_2$ is the wavenumber inside the cavity,
$\nu_2=1/(1+r_1r_2e^{2ik_2d_2})$, $\mathcal{R}_1=(r_1+r_2e^{2ik_2d_2})\nu_2$,
$\mathcal{R}_2=r_2$,
$\mathcal{T}_1=t_1\nu_2$, $\mathcal{T}_2=t_2$, $\mathcal{R}_1'=r_1'$,
$\mathcal{R}_2'=(r_2'+r_1'e^{2ik_2d_2})\nu_2$, $\mathcal{T}_1'=t_1'$,
and $\mathcal{T}_2'=t_2'\nu_2$ with
the conventional single interface
Fresnel reflection and transmission coefficients for left incidence
$r_i$ and $t_i$, $i\in\{1,2\}$, and right incidence $r_i'$ and $t_i'$, $i\in\{1,2\}$.
In contrast to the electric and magnetic field values where resonance effects can substantially
increase the magnitude of the field inside a resonator,
the photon-number values inside the cavity and at the outputs
in Eq.~\eqref{eq:photons}
are always between the input field photon numbers. This essentially ensures that
in global thermal equilibrium all the photon numbers are equal and no photon-number
accumulation can occur inside the cavity at the equilibrium state.

The total force on the medium block due to a light beam incident
from vacuum can be calculated as a difference of electromagnetic
pressures on both sides of the cavity multiplied with the area $S$.
The LDOSs on different sides of the cavity are equal and,
therefore, by using the spectral electromagnetic pressure in Eq.~\eqref{eq:pressure}
the total spectral force due to a light beam becomes
\begin{equation}
 \langle\hat F\rangle_\omega=\frac{\langle\hat n_1\rangle-\langle\hat n_3\rangle}{\langle\hat n_{1+}\rangle}\langle\hat F_0\rangle_\omega,
 \label{eq:totalforce}
\end{equation}
where $\langle\hat F_0\rangle_\omega$ is the spectral force in the case
of a perfect reflector in vacuum. When applying the propagating photon numbers
in Eq.~\eqref{eq:photons} and the identity $|\mathcal{R}_1|^2+|\mathcal{T}_1\mathcal{T}_2|^2=1$,
the force in Eq.~\eqref{eq:totalforce} becomes $\langle\hat F\rangle_\omega=|\mathcal{R}_1|^2\langle\hat F_0\rangle_\omega$.
The force is thus naturally proportional to the total power reflection coefficient of the structure
$|\mathcal{R}_1|^2$. If the medium block would be coated with anti-reflective coatings,
the total force would be zero meaning that the interface forces on the first and the
second interface cancel each other in accordance with the field-kinetic
one-photon model in Sec.~\ref{sec:fieldkineticmodel}. The
spectral interface forces are, in this case, explicitly
given by $\langle\hat F_1\rangle_\omega=(1-n)\langle\hat F_0\rangle_\omega$
and $\langle\hat F_2\rangle_\omega=(n-1)\langle\hat F_0\rangle_\omega$,
where $n=\sqrt{\varepsilon_2}$ is the refractive index of the medium block.

\section{Conclusions}
\label{sec:conclusions}

In conclusion, our analysis suggests that when a photon
enters the crystal its energy and momentum
will be shared by the crystal and the
propagating light wave or the polariton. As the ratio
of energy and momentum
of the polariton is different from that of light in vacuum,
the light wave can not be
covariantly described by a pure photon state that has no
rest mass. The covariance can be restored by assuming that
the polariton propagating in a crystal is a coupled
state of a photon and the induced dipoles in the
medium with a small but finite polariton rest mass.
In contrast to the previous interpretations that only
the Abraham momentum would obey the constant CEV motion
of an isolated body, we have shown that the constant CEV motion
can also be obeyed by the polariton with the Minkowski momentum.
We have also used the QFED formalism to study 
the steady-state interface forces in the
corresponding medium block geometry.

\begin{acknowledgments}
This work has in part been funded by the Academy of Finland and the Aalto Energy Efficiency Research Programme.
\end{acknowledgments}



\begin{thebibliography}{10}

\bibitem{Cho2010}
A.~Cho, ``Century-long debate over momentum of light resolved?,'' {\em
  Science}~{\bf 327}, p.~1067, 2010.

\bibitem{Leonhardt2006}
U.~Leonhardt, ``Momentum in an uncertain light,'' {\em Nature}~{\bf 444},
  pp.~823--824, 2006.

\bibitem{Barnett2010b}
S.~M. Barnett, ``Resolution of the {A}braham-{M}inkowski dilemma,'' {\em Phys.
  Rev. Lett.}~{\bf 104}, p.~070401, Feb 2010.

\bibitem{Barnett2010a}
S.~M. Barnett and R.~Loudon, ``The enigma of optical momentum in a medium,''
  {\em Phil. Trans. R. Soc. A}~{\bf 368}(1914), pp.~927--939, 2010.

\bibitem{Leonhardt2014}
U.~Leonhardt, ``Abraham and {M}inkowski momenta in the optically induced motion
  of fluids,'' {\em Phys. Rev. A}~{\bf 90}, p.~033801, Sep 2014.

\bibitem{Campbell2005}
G.~K. Campbell, A.~E. Leanhardt, J.~Mun, M.~Boyd, E.~W. Streed, W.~Ketterle,
  and D.~E. Pritchard, ``Photon recoil momentum in dispersive media,'' {\em
  Phys. Rev. Lett.}~{\bf 94}, p.~170403, May 2005.

\bibitem{Sapiro2009}
R.~E. Sapiro, R.~Zhang, and G.~Raithel, ``Atom interferometry using
  {K}apitza-{D}irac scattering in a magnetic trap,'' {\em Phys. Rev. A}~{\bf
  79}, p.~043630, Apr 2009.

\bibitem{Jones1954}
R.~V. Jones and J.~C.~S. Richards, ``The pressure of radiation in a refracting
  medium,'' {\em Proc. R. Soc. Lond. A}~{\bf 221}(1147), pp.~480--498, 1954.

\bibitem{Jones1978}
R.~V. Jones and B.~Leslie, ``The measurement of optical radiation pressure in
  dispersive media,'' {\em Proc. R. Soc. Lond. A}~{\bf 360}(1702),
  pp.~347--363, 1978.

\bibitem{Walker1975}
G.~B. Walker and D.~G. Lahoz, ``Experimental observation of {A}braham force in
  a dielectric,'' {\em Nature}~{\bf 253}, pp.~339--340, 1975.

\bibitem{She2008}
W.~She, J.~Yu, and R.~Feng, ``Observation of a push force on the end face of a
  nanometer silica filament exerted by outgoing light,'' {\em Phys. Rev.
  Lett.}~{\bf 101}, p.~243601, Dec 2008.

\bibitem{Zhang2015}
L.~Zhang, W.~She, N.~Peng, and U.~Leonhardt, ``Experimental evidence for
  {A}braham pressure of light,'' {\em New J. Phys.}~{\bf 17}(5), p.~053035,
  2015.

\bibitem{Schwartz2014}
M.~D. Schwartz, {\em Quantum field theory and the standard model}, Cambridge
  University Press, Cambridge, 2014.

\bibitem{Partanen2014a}
M.~Partanen, T.~H\"ayrynen, J.~Oksanen, and J.~Tulkki, ``Thermal balance and
  photon-number quantization in layered structures,'' {\em Phys. Rev. A}~{\bf
  89}, p.~033831, Mar 2014.

\bibitem{Partanen2014c}
M.~Partanen, T.~H\"ayrynen, J.~Oksanen, and J.~Tulkki, ``Unified
  position-dependent photon-number quantization in layered structures,'' {\em
  Phys. Rev. A}~{\bf 90}, p.~063804, Dec 2014.

\bibitem{Partanen2015a}
M.~Partanen, T.~H\"ayrynen, J.~Tulkki, and J.~Oksanen,
  ``Commutation-relation-preserving ladder operators for propagating optical
  fields in nonuniform lossy media,'' {\em Phys. Rev. A}~{\bf 92}, p.~033839,
  Sep 2015.

\bibitem{Feynman1964}
R.~Feynman, R.~Leighton, and M.~Sands, {\em The Feynman Lectures on Physics},
  Addison-Wesley, Massachusetts, 1964.

\bibitem{Partanen2014b}
M.~Partanen, T.~H{\"a}yrynen, J.~Oksanen, and J.~Tulkki, ``Position-dependent
  photon operators in the quantization of the electromagnetic field in
  dielectrics at local thermal equilibrium,'' in {\em Proc. SPIE 9136,
  Nonlinear Optics and Its Applications VIII; and Quantum Optics III},
  (91362B), SPIE, 2014.

\bibitem{Rodriguez2011}
A.~W. Rodriguez, F.~Capasso, and S.~G. Johnson, ``The {C}asimir effect in
  microstructured geometries,'' {\em Nature Photonics}~{\bf 5}, pp.~211--221,
  2011.

\bibitem{Antezza2008}
M.~Antezza, L.~P. Pitaevskii, S.~Stringari, and V.~B. Svetovoy,
  ``{C}asimir-{L}ifshitz force out of thermal equilibrium,'' {\em Phys. Rev.
  A}~{\bf 77}, p.~022901, Feb 2008.

\bibitem{Passante2007}
R.~Passante and S.~Spagnolo, ``Casimir-{P}older interatomic potential between
  two atoms at finite temperature and in the presence of boundary conditions,''
  {\em Phys. Rev. A}~{\bf 76}, p.~042112, Oct 2007.

\bibitem{Sushkov2011}
A.~O. Sushkov, W.~J. Kim, D.~A.~R. Dalvit, and S.~K. Lamoreaux, ``Observation
  of the thermal {C}asimir force,'' {\em Nature Physics}~{\bf 7}, pp.~230--233,
  2011.

\bibitem{Klimchitskaya2008}
G.~L. Klimchitskaya, U.~Mohideen, and V.~M. Mostepanenko, ``Thermal
  {C}asimir-{P}older force between an atom and a dielectric plate:
  thermodynamics and experiment,'' {\em J. Phys. A}~{\bf 41}(43), p.~432001,
  2008.

\end{thebibliography}
\end{document}